\documentclass[twocolumn,showpacs,preprintnumbers,amsmath,amssymb]{revtex4}
\usepackage{graphicx}% Include figure files
\usepackage{dcolumn}% Align table columns on decimal point
\usepackage{bm}% bold math
\def\jpsi{{J/\psi}}

\def\be{\begin{equation}}
\def\ee{\end{equation}}
\def\bea{\begin{eqnarray}}
\def\eea{\end{eqnarray}}
\def\NO{\nonumber}
\def\gev{\mathrm{~GeV}}
\def\fb{\mathrm{~fb}}
\def\dfrac{\displaystyle\frac}
\def\co{{\cal O}}
\def\a{\alpha}
\def\b{\beta}

\def\e{\epsilon}
\def\g{\gamma}
\def\s{\sigma}
\begin{document}
%\preprint{APS/123-QED}

\title{QCD corrections to $\jpsi$ plus $\eta_c$ production in $e^+e^-$ annihilation at $\sqrt{s}=10.6 \gev$}% Force line breaks with \\

\author{Bin Gong and Jian-Xiong Wang}%
\affiliation{
Institute of High Energy Physics, Chinese Academy of Sciences, P.O. Box 918(4), Beijing, 100049, China.
}%
\date{\today}% It is always \today, today,

\begin{abstract}
Next-to-Leading-Order(NLO) QCD corrections to $\jpsi$ plus $\eta_c$ production in $e^+e^-$ annihilation at $\sqrt{s}=10.6 \gev$ is calculated in this paper, and an analytic result is obtained. By choosing proper physical parameters, a K factor (ratio of NLO to LO) of about 2, which is in agreement with the result in Ref.~\cite{Zhang:2005ch}, is obtained. 
%Our results show that the Next-Next-to-Leading-Order(NNLO) corrections might be quite large.
The plot of the K-factor vs the center-of-mass energy $\sqrt{s}$  shows that it is more difficult to obtain a convergent result from the perturbative QCD without resummation of $\ln(s/m_c^2)$ terms as the $\sqrt{s}$ becomes larger.   
\end{abstract}

%\keywords{Suggested keywords}%Use showqkeys class option if keyword
\maketitle

\section{Introduction}
Perturbative quantum chromodynamics calculations are essential in the effort to describe large momentum transfer processes. To apply it to heavy quarkonium physics, the nonrelativistic QCD (NRQCD) factorization approach\cite{Bodwin:1994jh} has been introduced. It allows consistent theoretical prediction to be made and to be improved perturbatively in the QCD coupling constant $\a_s$ and the heavy-quark relative velocity $v$. However, the $\jpsi$ polarization measurement at Fermilab Tevatron in proton-antiproton collisions~\cite{Abulencia:2007us} and $\jpsi$ production in B-factories ~\cite{Abe:2002rb,Pakhlov:2004au,Aubert:2005tj} have shown that the leading order (LO) theoretical predictions in NRQCD could not match the experimental results. The large discrepancy was found in the double charm production in $e^+e^-$ annihilation at B factories. The exclusive production cross section of double charmonium in $e^+e^-\rightarrow \jpsi\eta_c$ at $\sqrt{s}=10.6$ GeV measured by Belle \cite{Abe:2002rb,Pakhlov:2004au} is 
$\s[\jpsi+\eta_c] \times B^{\eta_c}[\geq2] = (25.6\pm2.8\pm3.4)\fb$ 
and by BABAR \cite{Aubert:2005tj} is
$\s[\jpsi+\eta_c] \times B^{\eta_c}[\geq2] = (17.6\pm2.8^{+1.5}_{-2.1})\fb$,
where $B^{\eta_c}[\geq2]$ denotes the branching fraction for the $\eta_c$ decaying into at least two charged tracks. 
Meanwhile, the NRQCD LO theoretical predictions in the QCD coupling constant $\a_s$ and the charm-quark relative velocity $v$,  given by Braaten and Lee \cite{Braaten:2002fi}, Liu, He and Chao \cite{Liu:2002wq}, and  Hagiwara, Kou and Qiao \cite{Hagiwara:2003cw}
are about $2.3 \sim 5.5 \fb$, 
which is an order of magnitude smaller than the experimental results. Such a large discrepancy between experimental results and theoretical predictions brings a challenge to the current understanding of charmonium production based on NRQCD. Many studies have been performed in order to resolve the problem. 
Braaten and Lee \cite{Braaten:2002fi} have shown that the relativistic corrections would increase the cross section by a factor of about 2, which boost the cross section to $7.4 \fb$. 
And the NLO QCD correction of the process has been studied by Zhang, Gao and Chao \cite{Zhang:2005ch}, which can enhance the cross section with a $K$ factor (the ratio of NLO to LO) of about 2 and reduce the large discrepancy.
Again the relativistic corrections have been studied by Bodwin, Kang, Kim, Lee and Yu \cite{Bodwin:2006ke} and by He, Fan and Chao \cite{He:2007te}, which are significant, and when combined with the NLO QCD corrections, may resolve the large discrepancy. In Ref.~\cite{Jia:2007hy}, $\Upsilon(4s)\rightarrow \jpsi +\eta_c$ was considered by Jia, but it's contribution is small.  Ma and Si \cite{Ma:2004qf} treated the process by using light-cone method.
A similar treatment was performed by Bondar and Chernyad \cite{bondar:2005} and Bodwin, Kang and Lee \cite{Bodwin:2006dm}.
More detailed treatment, such as including the resummation of a class of relativistic correction, has been take into consideration by Bodwin and Lee and Yu \cite{Bodwin:2007ga}.

Since the calculation of NLO QCD correction for this process is quite complicate and plays a very important role to explain the experimental data, in this paper we performed an independent calculation by using the package Feynman Diagram Calculation (FDC) \cite{FDC} with one-loop part built in and obtained analytic result. The numerical result is in agreement with the previous result in Ref.~\cite{Zhang:2005ch}.

This paper is organized as follows. In Sec. II, we give the LO cross section for the process. 
The calculation of NLO QCD corrections are described in Sec. III. In Sec. IV, numerical results are presented. The conclusion
and discussion are given in Sec. V. In the Appendixes, some useful details are presented.
\section{The LO cross section}
\begin{figure}
\center{
\includegraphics*[scale=0.6]{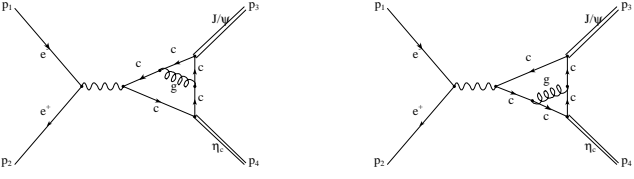}% Here is how to import EPS art
\caption {\label{fig:LO}Feynman diagrams for LO. }}
\end{figure}

There are four Feynman diagrams for this order, and two of which are shown in Fig.~\ref{fig:LO}, while the other two can be obtained by reversing the arrows of the quark lines. Momenta for the involved particles are labeled as 
\be
e^-(p_1)+ e^+(p_2) \rightarrow \jpsi(p_3) + \eta_c(p_4). 
\ee
In the nonrelativistic limit, we can use the NRQCD factorization formalism and obtain the square of the scattering amplitude as
\bea
|M_{LO}|^2&=&\dfrac{2^{14}\pi^2\a^2\a_s^2e_c^2|R_s^\jpsi(0)|^2|R_s^{\eta_c}(0)|^2 }{9m_c^6s^5} \NO\\
&&\times\biggl(2-4s+s^2-4t+2st+2t^2\biggr),
\eea
with
\be
s=\dfrac{(p_1+p_2)^2}{4m_c^2},\quad t=\dfrac{(p_1-p_3)^2}{4m_c^2}, 
\label{eqn:s}
\ee
where $e_c=\frac{2}{3}$ is the electric charge of the charm quark. $R_s^{\jpsi}(0)$ and $R_s^{\eta_c}(0)$ are the radial wave function at the origin of $\jpsi$ and $\eta_c$. Notice that $s$ in Eq.~(\ref{eqn:s}) is used from now on.
After the integration of phase space, the total cross section is
\be
\s^{(0)}=\dfrac{128\pi\a^2\a_s^2e_c^2|R_s^\jpsi(0)|^2|R_s^{\eta_c}(0)|^2 (s-4)^{\frac{3}{2}}}{27m_c^8s^{\frac{11}{2}}} .
\ee 
\section{The NLO cross section}
\begin{figure}
\center{
\includegraphics*[scale=0.5]{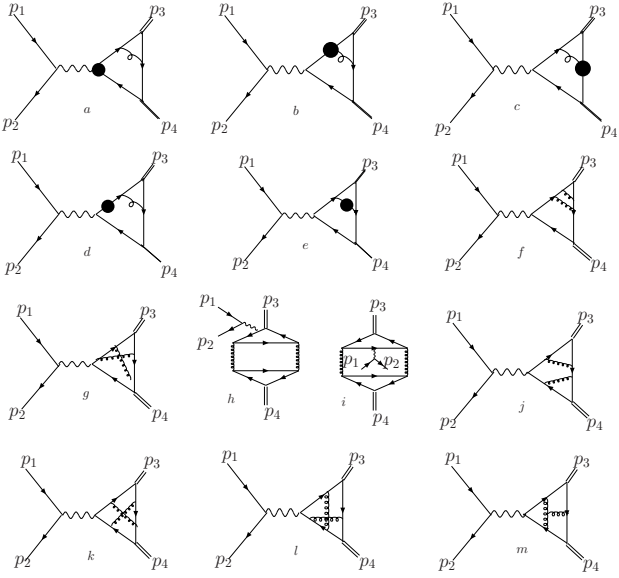}% Here is how to import EPS art
\caption {\label{fig:NLO}All Feynman diagrams for NLO are divided into thirteen groups. (a) includes the photon-quark vertex counter-term and corresponding loop diagrams; (b) and (c) are the gluon-quark vertex counter-term and corresponding loop diagrams; (d) and (e) denote the counter-term and corresponding loop diagrams for the quark and gluon self-energy; (f) and (j) are diagrams that contain Coulomb singularity. Other diagrams can be obtained by reversing the arrows of quark lines and/or change the places of $\jpsi$ and $\eta_c$. But notice that we can't change the places of $\jpsi$ and $\eta_c$ in group (h) and (i).}}
\end{figure}

Since there is no $\co(\alpha_s)$ real process in NLO, we only need to calculate virtual corrections.  Dimensional regularization has been
adopted for isolating the ultraviolet(UV) and infrared(IR) singularities.
UV divergences from self-energy and triangle diagrams are canceled upon the renormalization of the QCD gauge coupling constant, the charm quark mass and field, and the gluon field. A similar renormalization scheme is chosen as in ref.~\cite{Klasen:2004tz} except that both light quarks and charm quark are included in the quark loop to obtain the renormalization constants. The renormalization constants of the charm quark mass $Z_m$ and field $Z_2$, and the gluon field $Z_3$ are defined in the on-mass-shell(OS) scheme while that of the QCD gauge coupling $Z_g$ is defined in the modified-minimal-subtraction($\overline{\mathrm{MS}}$) scheme:
\bea
\delta Z_m^{OS}&=&-3C_F\dfrac{\alpha_s}{4\pi}\left[\dfrac{1}{\e_{UV}} -\gamma_E +\ln\dfrac{4\pi \mu^2}{m_c^2} +\frac{4}{3} +\co(\e)\right] , \NO\\
\delta Z_2^{OS}&=&-C_F\dfrac{\alpha_s}{4\pi}\NO\\&&\times\biggl[\dfrac{1}{\e_{UV}} +\dfrac{2}{\e_{IR}} -3\gamma_E +3\ln\dfrac{4\pi \mu^2}{m_c^2} +4 +\co(\e)\biggr] , \NO\\
\delta Z_3^{OS}&=&\dfrac{\alpha_s}{4\pi}\biggl[(\beta'_0-2C_A)\left(\dfrac{1}{\e_{UV}} -\dfrac{1}{\e_{IR}}\right)
\NO\\&&
-\dfrac{4}{3}T_F\left(\dfrac{1}{\e_{UV}} -\gamma_E +\ln\dfrac{4\pi \mu^2}{m_c^2}\right) +\co(\e)\biggr] , \NO\\
\delta Z_g^{\overline{\mathrm{MS}}}&=&-\dfrac{\beta_0}{2}\dfrac{\alpha_s}{4\pi}\left[\dfrac{1}{\e_{UV}} -\gamma_E +\ln(4\pi) +\co(\e)\right] . 
\eea
where $\g_E$ is Euler's constant, $\b_0=\frac{11}{3}C_A-\frac{4}{3}T_Fn_f$ is the one-loop coefficient of the QCD beta function and $n_f$ is the number of active quark flavors. There are three massless light quarks $u, d, s$, and one heavy quark $c$, so $n_f$=4. In $SU(3)_c$, color factors are given by $T_F=\frac{1}{2}, C_F=\frac{4}{3}, C_A=3$. And $\b'_0\equiv\b_0+(4/3)T_F=(11/3)C_A-(4/3)T_Fn_{lf}$ where $n_{lf}\equiv n_f-1=3$ is the number of light quarks flavors. Actually in the NLO total amplitude level, the terms proportion to $\delta {Z_3}^{OS}$  cancel each other, thus the result is independent of renormalization scheme of the gluon field.

After having fixed our renormalization scheme and omitting diagrams that do not contribute, including counter-term diagrams, there are 80 NLO diagrams remained, which are shown in Fig.~\ref{fig:NLO}. It is divided into 13 groups. Diagrams of group $(f)$ and $(j)$ that have a virtual gluon line connected with the quark pair in a meson lead to Coulomb singularity $\sim \pi^2/v$, which can be isolated by introducing a small relative velocity $v=|\vec{p}_{c}-\vec{p}_{\bar{c}}|$. The corresponding contribution is also of $\co(\a_s)$ and can be taken into the $c\bar{c}$ wave function renormalization\cite{Kramer:1995nb} as: 
\bea
\s&=&|R_s(0)|^2\hat{\s}^{(0)}\left( 1 +\dfrac{\a_s}{\pi}C_F\dfrac{\pi^2}{v} + \dfrac{\a_s}{\pi}C +\co(\a_s^2)\right)  \NO\\
%&=&|R_s(0)|^2\left( 1 +\dfrac{\a_s}{\pi}C_F\dfrac{\pi^2}{v}\right) \hat{\s}^{(0)} 
%\left[1+ \dfrac{\a_s}{\pi}C +\co(\a_s^2)\right] \NO\\
&\Rightarrow&|R^{ren}_s(0)|^2 \hat{\s}^{(0)} \left[1+ \dfrac{\a_s}{\pi}C +\co(\a_s^2)\right].
\eea
A factor of 2 should be used since there are two bound states. 
After adding contributions from all the diagrams together, all the IR-divergent terms are canceled and the total scattering amplitude is obtained as
\bea
M_{NLO}+M_{LO}&=&M_{LO}\left\{1+\dfrac{\a_s(\mu)}{2\pi}\left[\dfrac{8}{3}\dfrac{\pi^2}{v} -\beta_0\ln\dfrac{2m_c}{\mu} 
\right.\right.\NO\\&&\left.\left.
+\dfrac{K_1(s)}{6}+i\pi\dfrac{K_2(s)}{6}\right]\right\},
\label{eqn:M_NLO}
\eea
with $K_1(s)$ and $K_2(s)$ given by Eqs.~(\ref{eqn:k1}) and (\ref{eqn:k2}).

Meanwhile $\a_s$ should be obtained from two-loop formula as
\be
\dfrac{\a_s(\mu)}{4\pi} =\dfrac{1}{\b_0 \ln(\mu^2/\Lambda_{QCD}^2)} - \dfrac{\b_1 \ln \ln(\mu^2/\Lambda_{QCD}^2)}{\b_0^3 \ln^2(\mu^2/\Lambda_{QCD}^2)},
\ee
where 
%\be
%\b_1=\dfrac{34}{3}C_A^2-4\left(C_F+\dfrac{5}{3}C_A\right)T_Fn_f
%\ee
$\b_1=34C_A^2/3-4(C_F+5C_A/3)T_F n_f$
is two-loop coefficient of the QCD beta function. From Eq.~(\ref{eqn:M_NLO}) the total cross section at NLO is: 
\be
\s_{NLO}=\s^{(0)}\left\{1+\dfrac{\a_s(\mu)}{\pi}\left[-\beta_0\ln\dfrac{2m_c}{\mu} +\dfrac{K_1(s)}{6}\right]\right\}.
\ee

%By directly performing the square of NLO scattering amplitude, a gauge invariant, UV and IR divergence free NNLO part could be included. It gives the total cross section as:
%\be
%\s_{NLO^+}=\s^{(0)}\left\{1+\dfrac{\a_s(\mu)}{\pi}\left[-\beta_0\ln\dfrac{2m_c}{\mu} +\dfrac{K_1(s)}{6}\right]+ \dfrac{\a_s^2(\mu)}{4\pi^2}\left[\biggl(-\beta_0\ln\dfrac{2m_c}{\mu} +\dfrac{K_1(s)}{6}\biggr)^2+\biggl(\dfrac{\pi K_2(s)}{6}\biggr)^2\right]
%\right\}\label{eq:one}.  
%\ee
%$\s_{NLO^+}$ could be used as a kind of uncertainty estimation to the NNLO correction.
%
\section{Numerical result}
Up to NLO, the value of the wave function at the origin of $\jpsi$ is related to the leptonic decay widths as:
\be
\Gamma_{ee}=\left(1-\dfrac{16}{3}\dfrac{\a_s}{\pi}\right) \dfrac{4\alpha^2e_c^2}{M_{\jpsi}^2}|R_s^\jpsi(0)|^2 ,
\ee
and according to ref.~\cite{Bodwin:1994jh}, we can set $R_s^{\eta_c}(0)=R_s^\jpsi(0)=R_s(0)$. If we choose $|R_s(0)|^2=0.978 \gev^3$ and $\Lambda^{(4)}_{\overline{\mathrm{MS}}}=0.338 \gev$,
then we get the numerical result as shown in TABLE.~\ref{table:result}, which is consistent with the result in ref.~\cite{Zhang:2005ch}. 
%The results of $\s_{NLO^+}$ show that the NNLO correction might be quite large.  
\begin{table}[htbp]
\begin{center}
\begin{tabular}{|c|c|c|c|c|c|}
\hline\hline
$m_c$(GeV)	&$\mu$	&$\a_s(\mu)$	&$\s_{LO}$(fb)&$\s_{NLO}$(fb)	&$\s_{NLO}/\s_{LO}$\\
\hline
1.5&$m_c$&0.369&16.09&27.51&1.710\\
\hline
1.5&2$m_c$&0.259&7.94&15.68 &1.975\\
\hline
1.5&$\sqrt{s_0}/2$&0.211&5.27&11.14&2.114\\
\hline
1.4&$m_c$&0.386&19.28&34.92 &1.811\\
\hline
1.4&2$m_c$&0.267&9.19&18.84&2.050 \\
\hline
1.4&$\sqrt{s_0}/2$&0.211&5.76&12.61&2.189\\
\hline\hline
\end{tabular}
\caption{Cross sections with different charm quark mass $m_c$ and renormalization scale $\mu$. $\sqrt{s_0}=10.6 \gev$ is the center-of-mass energy.}
\label{table:result}
\end{center}
\end{table} 
\begin{figure}
\center{
\includegraphics*[scale=0.45]{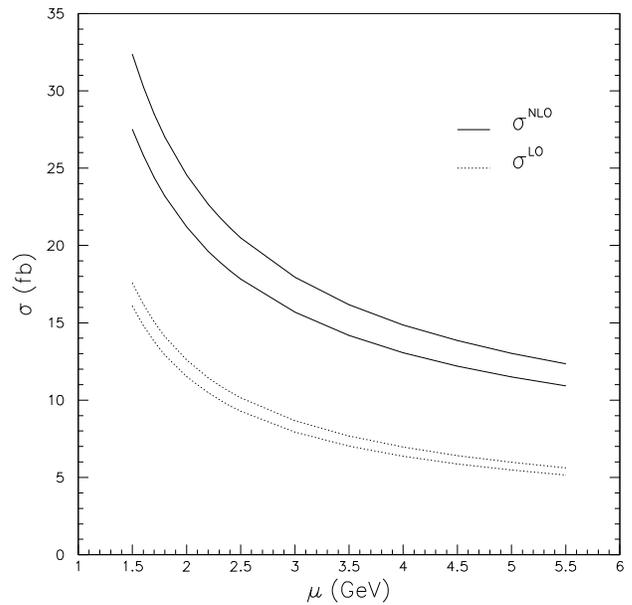}% Here is how to import EPS art
\caption {\label{fig:result}Cross sections as function of the renormalization scale $\mu$ with $|R_s(0)|^2=0.978\gev^3 , \Lambda=0.338\gev$ and center-of-mass energy $10.6\gev$. The charm quark mass is chosen as 1.4 GeV (upper curves) and 1.5 GeV (lower curves).
%while LO results are represented by dot lines and NLO ones are represented by solid lines. 
%and dash lines represent  $NLO^+$ results. 
%Upper curves are for  $m_c=1.4\gev$ while the lower ones are for $m_c=1.5\gev$.
}}
\end{figure}
\begin{figure}
\center{
\includegraphics*[scale=0.45]{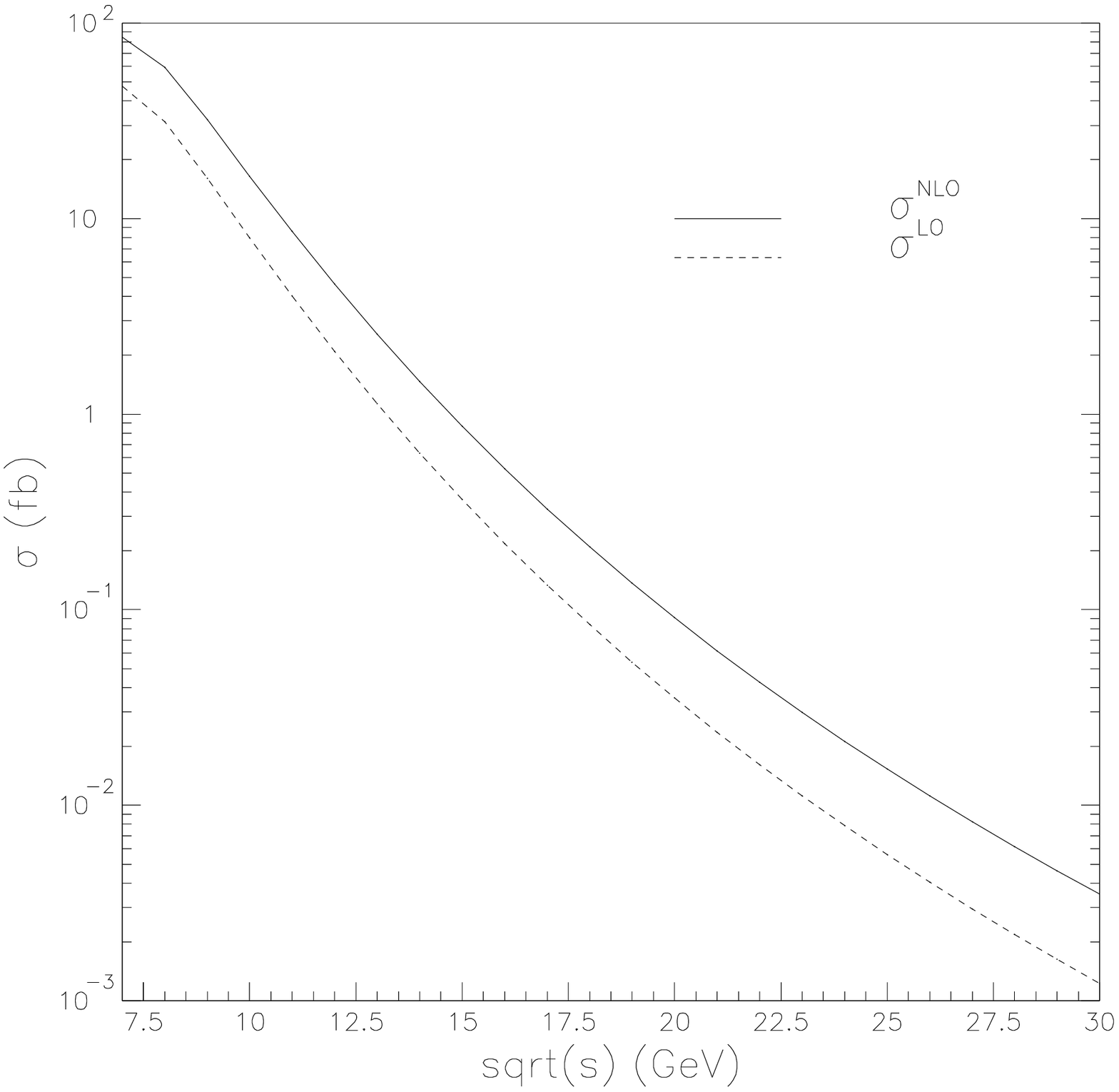}% Here is how to import EPS art
\caption {\label{fig:result2}Cross sections as function of the center-of-mass energy with $|R_s(0)|^2=0.978\gev^3$ and $\Lambda=0.338 \gev$. Renormalization scale $\mu$ is set at half of the center-of-mass energy and $m_c=1.5\gev$.}}
\end{figure}
\begin{figure}
\center{
\includegraphics*[scale=0.45]{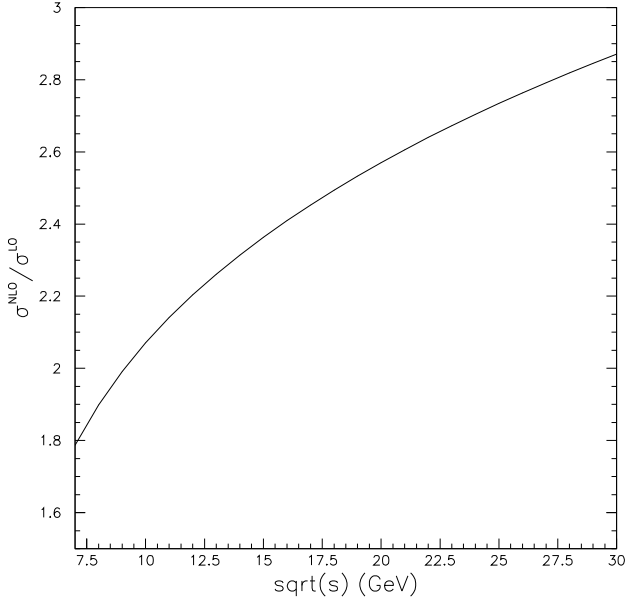}% Here is how to import EPS art
\caption {\label{fig:k}K factor as function of the center-of-mass energy with $|R_s(0)|^2=0.978\gev^3$ and $\Lambda=0.338\gev$. Renormalization scale $\mu$ is set at half of the center-of-mass energy and $m_c=1.5\gev$.}}
\end{figure}
\section{Conclusion}
We calculated the NLO QCD correction of $\jpsi$ plus $\eta_c$ production in $e^+e^-$ annihilation at center-of-mass energy $10.6$ GeV. The method of dimensional regularization is taken to deal with the UV and IR singularities, and the Coulomb singularity is isolated by a small relative velocity $v$ between the charm quark pair in the meson and absorbed into the $c\bar{c}$ bound state wave function. 
After taking all one-loop diagrams into account, an analytic finite result is obtained. By choosing proper physical parameters, we get a K factor (ratio of NLO to LO) of about 2, which is consistent with ref.~\cite{Zhang:2005ch}. It decreases the great discrepancy between theory and experiment. 
%Furthermore, if we include part of the NNLO correction as $\s_{NLO^+}$ in Eq.~(\ref{eq:one}), the cross section will be enhanced for about 40-50\% extra. 
From Fig.~\ref{fig:result}, it could be found that the dependence on the renormalization scale $\mu$ has not been improved in NLO calculation. The plot of the total cross section vs the center-of-mass energy of $e^+e^-$ in Fig.~\ref{fig:result2} behaves as expected. But the plot of the K-factor vs the center-of-mass energy of $e^+e^-$ in Fig.~\ref{fig:k} shows that it is more difficult to obtain the convergent result from the perturbative QCD without resummation of $\displaystyle{\ln{\frac{s}{m_c^2}}}$ terms as the center-of-mass energy of $e^+e^-$ becomes larger.   
\begin{acknowledgments}
This work was supported by the National Natural Science Foundation of China (No.~10475083).
\end{acknowledgments}
\appendix
\section{The definition of $K_1$ and $K_2$}
In this section, the definition $K_1(s)$ and $K_2(s)$ used in the Eq.~(\ref{eqn:M_NLO}) are presented.
\begin{widetext}
\bea
K_1(s)&=& 
%{\displaystyle{ 4z_{1} \over s}} 
{\displaystyle{ {-24-79s-68s^2} \over 3s(2s+1)}} +f_{1}{\displaystyle{s^2-3s+16 \over 2(s-4)}} 
%+f_{2}{\displaystyle{3s \over 2(s-2)}} 
-f_{2}{\displaystyle{3s \over 2(s-2)}} 
+f_{3}{\displaystyle{-34s^3+193s^2-342s+160 \over4(s-4)(s-2)}}
\NO\\[3mm]&&
 +f_{4}{\displaystyle{-41s^2-194s+64 \over 32(s-4)}} +f_{5}{\displaystyle{8s^2-21s-8 \over 2(s-4)}} 
+f_{6}{\displaystyle{65s^2-302s+64 \over 32(s-4)}} 
+f_{7}{\displaystyle{-3s^2-4s \over 2(s-4)}} +a_{1}\biggl[{\displaystyle{ 4z_{1}(8-7s) \over s^2(s-4)}} 
\NO\\[3mm]&&
+{\displaystyle{ -64s^4+406s^3+11s^2-335s-120 \over (s-4)(2s+1)^2}}\biggr] +a_{6}\biggl[{\displaystyle{ 2z_{1}(8-7s) \over s^2(s-4)}} +{\displaystyle{
 z_{2}(-130+19s+8s^2) \over s(s-4)}} 
\NO\\[3mm]&&
+{\displaystyle{ -32s^4-110s^3+639s^2+696s+172 \over (s-4)(2s+1)^2}}\biggr] %\NO\\[3mm]&&
+a_{7}{\displaystyle{4z_{1}(7s-8) \over s^2(s-4)}} 
+a_{9}{\displaystyle{2z_{2}(-8s^2-19s+130) \over s(s-4)}}
%\\[3mm]
\label{eqn:k1},\\[3mm]
%\eea
%\bea
K_2(s)&=&
2z_{1}{\displaystyle{-7s+8 \over (s-4)s^2}}
+z_{2}{\displaystyle{-8s^2-19s+130 \over (s-4)s}}
+{\displaystyle{32s^4+110s^3-639s^2-696s-172 \over (s-4)(2s+1)^2}} 
+z_{1}{\displaystyle{(65s^2-302s+64) \over 4s(s-4)^2}}a_{13} 
\NO\\[3mm]&&
+z_{1}{\displaystyle{(21s^3-64s^2-1396s+2688) \over 8(s-2)s(s-4)^2}}a_{1}  
+z_{1}{\displaystyle{(73s^3-96s^2+444s-896) \over 8(s-2)s(s-4)^2}}a_{3}  
 \NO\\[3mm] &&
-z_{1}{\displaystyle{(193s^3-1292s^2+2548s-1280) \over 8(s-2)s(s-4)^2}}a_{5} +z_{1}{\displaystyle{(79s^2-82s+64) \over 8s(s-4)^2}}a_{6} 
-4z_{1}{\displaystyle{(8s^2-21s-8) \over s(s-4)^2}}a_{7}
  \NO\\[3mm]&&
 +z_{1}{\displaystyle{(34s^3-193s^2+342s-160) \over 2(s-2)s(s-4)^2}}(a_{10}-a_{8})  %\NO\\[3mm]&&
+z_{1}{\displaystyle{(183s^3-1624s^2+3316s-1408) \over 8(s-2)s(s-4)^2}}a_{12} .
\label{eqn:k2}
\eea
%\end{widetext}
Where all the variables used in $K_1(s)$ and $K_2(s)$ are defined as:
\be
z_{1}= \sqrt{s^2-4s}, \quad z_{2}= \sqrt{s^2-s},
\ee
\bea
f_{1}&=&{\displaystyle{4z_{1} \over s^2-4s}} (-2a_{1}^2-a_{1}a_{2}-a_{1}a_{3}+a_{1}a_{4}+2a_{1}a_{6}
+4a_{1}a_{7}-a_{2}a_{3}+a_{3}^2+a_{3}a_{4}-a_{3}a_{5}\NO\\
&&-2a_{3}a_{6}+a_{5}a_{6}+a_{6}^2-2a_{6}a_{7}-l_{1} +l_{2}-l_{3}-l_{4}+l_{5}-l_{6}+l_{7}+2l_{8}) ,\NO\\
f_{2}&=&{\displaystyle{2z_{1} \over s^2-4s}}(6a_{1}^2-2a_{1}a_{3}+a_{1}a_{5}-4a_{1}a_{7}-2a_{3}a_{6}+a_{5}a_{6} -2a_{6}^2+4a_{6}a_{7}+l_{10}-l_{11}
%\NO\\&&
-l_{12} +l_{13}+2l_{14}-l_{9}),\NO\\
f_{3}&=&{\displaystyle{2z_{1} \over s^2-4s}}(2a_{1}a_{12}-a_{1}a_{5}+a_{10}a_{6}-2a_{10}a_{9}-2a_{11}a_{12}  -2a_{12}^2+2a_{12}a_{5}+2a_{12}a_{6}\NO\\
&&+2a_{12}a_{7}-a_{5}a_{6}-a_{6}a_{8}+2a_{8}
a_{9}-2l_{15}+l_{16}+l_{17}+l_{18}-l_{19}-l_{20}+l_{21}-l_{22}),\NO\\
f_{4}&=&{\displaystyle{4z_{1} \over s^2-4s}}(-2a_{1}a_{12}+a_{1}a_{5} -a_{1}a_{6}-a_{12}a_{6}+a_{3}a_{6}-2l_{15}+l_{23}+l_{24}-l_{25}-l_{26} +l_{27}),\NO\\
f_{5}&=&{\displaystyle{4z_{1} \over s^2-4s}}(-2a_{1}a_{6}-a_{6}^2+2a_{6}a_{7}-2l_{15}+l_{26}-l_{27}+2l_{28}),\NO \\
f_{6}&=&{\displaystyle{4z_{1} \over s^2-4s}}(-2a_{1}a_{13}+a_{1}a_{5}+2a_{1}a_{6}+2a_{1}a_{7}
-2a_{13}a_{6}+a_{5}a_{6}+a_{6}^2-l_{16}-l_{21}+l_{22}
% \NO\\&&
-l_{26}+l_{27}+2l_{28}),\NO\\
f_{7}&=&{\displaystyle{2z_{1} \over s^2-4s}}(-2a_{1}a_{14}+2a_{1}a_{15}-a_{1}a_{5} +a_{12}^2 -a_{12}a_{14}+a_{12}a_{15}-a_{12}a_{5}-2a_{12}a_{6} \NO\\
&&+a_{5}a_{6}-2l_{15}-l_{16}-l_{21}+l_{22}+4l_{28}) ,
\eea
\end{widetext}
\bea
a_{1}&=&\mathrm{aln}(2),\NO\\
a_{2}&=&\mathrm{aln}(s^2-sz_{1}-2z_{1}),\NO\\
a_{3}&=&\mathrm{aln}(s-z_{1}+2),\NO\\
a_{4}&=&\mathrm{aln}(s^2-sz_{1}+2s-z_{1}),\NO\\
a_{5}&=&\mathrm{aln}(2s+1),\NO\\
a_{6}&=&\mathrm{aln}(s),\NO\\
a_{7}&=&\mathrm{aln}(s+z_{1}),\NO\\
a_{8}&=&\mathrm{aln}(3s^2+sz_{1}-sz_{2}-6s-3z_{1}z_{2}-2z_{1}+4z_{2}),\NO\\
a_{9}&=&\mathrm{aln}(s-z_{2}),\NO\\
a_{10}&=&\mathrm{aln}(3s^2+sz_{1}+sz_{2}-6s+3z_{1}z_{2}-2z_{1}-4z_{2}),\NO\\
a_{11}&=&\mathrm{aln}(2s^2+2sz_{1}-5s-z_{1}),\NO\\
a_{12}&=&\mathrm{aln}(s-z_{1}-1),\NO\\
a_{13}&=&\mathrm{aln}(3s-z_{1}),\NO\\
a_{14}&=&\mathrm{aln}(s^2-sz_{1}-3s+z_{1}),\NO\\
a_{15}&=&\mathrm{aln}(s^2-sz_{1}-s-z_{1}),\NO\\
\eea
and
\bea
l_{1}&=&{\mathrm{Li}_2^r}({s^2+sz_{1}-4s+2z_{1} \over 8s^2+4s}),\NO\\
l_{2}&=&{\mathrm{Li}_2^r}({-s^2-sz_{1}+4s-2z_{1} \over 4s+2}),\NO\\
l_{3}&=&{\mathrm{Li}_2^r}({-s^2+sz_{1}+4s+2z_{1} \over 4s}),\NO\\
l_{4}&=&{\mathrm{Li}_2^r}({-s-z_{1}+4 \over 8}),\NO\\
l_{5}&=&{\mathrm{Li}_2^r}({-s+z_{1}+4 \over 8}),\NO\\
l_{6}&=&{\mathrm{Li}_2^r}({s-z_{1}-4 \over 2s}),\NO\\
l_{7}&=&{\mathrm{Li}_2^r}({s+z_{1}-4 \over 2s}),\NO%\\
\eea
\bea
l_{8}&=&{\mathrm{Li}_2^r}({z_{1} \over 2s}),\NO\\
l_{9}&=&{\mathrm{Li}_2^r}({s^3+s^2z_{1}-4s^2-2sz_{1}-2z_{1} \over 8s+4}),\NO\\
l_{10}&=&{\mathrm{Li}_2^r}({-s^4-s^3z_{1}+4s^3+2s^2z_{1}+2sz_{1} \over 4s+2}),
\NO\\
l_{11}&=&{\mathrm{Li}_2^r}({-s^3+s^2z_{1}+4s^2-2sz_{1}-2z_{1} \over 4}),\NO\\
l_{12}&=&{\mathrm{Li}_2^r}({-s^2-sz_{1}+4s+2z_{1} \over 2}),\NO\\
l_{13}&=&{\mathrm{Li}_2^r}({-s^2+sz_{1}+4s-2z_{1} \over 2}),\NO\\
l_{14}&=&{\mathrm{Li}_2^r}({z_{1} \over 2}),\NO\\
l_{15}&=&{\mathrm{Li}_2^r}({-z_{1} \over s}),\NO\\
l_{16}&=&{\mathrm{Li}_2^r}({s^2+sz_{1}-4s-z_{1} \over 2s^2+s}),\NO\\
l_{21}&=&{\mathrm{Li}_2^r}({-s^2+sz_{1}+4s-z_{1} \over s}),\NO\\
l_{22}&=&{\mathrm{Li}_2^r}({-2s^2-2sz_{1}+8s+2z_{1} \over 2s+1}),\NO\\
l_{23}&=&{\mathrm{Li}_2^r}({s^2-sz_{1}-4s+z_{1} \over 2s^2+s}),\NO\\
l_{24}&=&{\mathrm{Li}_2^r}({-s^2-sz_{1}+4s+z_{1} \over s}),\NO\\
l_{25}&=&{\mathrm{Li}_2^r}({-2s^2+2sz_{1}+8s-2z_{1} \over 2s+1}),\NO\\
l_{26}&=&{\mathrm{Li}_2^r}({-s-z_{1}+4 \over 2}),\NO\\
l_{27}&=&{\mathrm{Li}_2^r}({-s+z_{1}+4 \over 2}),\NO\\
l_{28}&=&{\mathrm{Li}_2^r}({z_{1} \over s}), \NO
\eea
\begin{widetext}
\bea
l_{17}&=&{\mathrm{Li}_2^r}({-s^3-s^2z_{1}-s^2z_{2}+5s^2-sz_{1}z_{2}+2sz_{1}+4sz_{2}-4s+2
z_{1}z_{2} \over s^2}),\NO\\
l_{18}&=&{\mathrm{Li}_2^r}({-s^3-s^2z_{1}+s^2z_{2}+5s^2+sz_{1}z_{2}+2sz_{1}-4sz_{2}-4s-2
z_{1}z_{2} \over s^2}),\NO\\
l_{19}&=&{\mathrm{Li}_2^r}({-s^3+s^2z_{1}-s^2z_{2}+5s^2+sz_{1}z_{2}-2sz_{1}+4sz_{2}-4s-2
z_{1}z_{2} \over s^2}),\NO\\
l_{20}&=&{\mathrm{Li}_2^r}({-s^3+s^2z_{1}+s^2z_{2}+5s^2-sz_{1}z_{2}-2sz_{1}-4sz_{2}-4s+2
z_{1}z_{2} \over s^2}).
\eea
In above expression, $\mathrm{aln}(x)=\ln|x|$ and $\mathrm{Li}_2^r(x)=\mathrm{Re}[\mathrm{Li_2}(x)]$.
\end{widetext}
\section{The results for all the scalar integrals}
In this section, we present the results of scalar integrals. Functions $A$ to $E$ denote one to five point scalar integrals and variables of the functions are written as $T(p_0,m_0,\dots,p_n,m_n)$ where $p_i$ and $m_i$ denote parameters of the nth propagator $N_i=(q+p_i)^2-m_i^2+i\e$. $q$ is the loop momentum and a factor of 
\be
C_\e=\dfrac{i}{16\pi^2}e^{-\e\gamma_E}\left(\dfrac{4\pi\mu^2}{4m_c^2}\right) 
\ee
is taken away from all the scalar integrals. 

We have developed a full series of methods in calculating tensor and scalar integrals with dimensional regularization 
and realized it in FDC\cite{FDC}. A paper about these methods are in preparation. All the scalar integrals are calculated 
analytically by using FDC and the results are shown in the following.
\begin{description}
\item One point:
\bea
A(0,{m_{c}})&=&4F_{14}{m_{c}}^2+{\frac{1}{\varepsilon_{UV}}}{m_{c}}^2 
\eea
\item Two point:
\bea
B(0,0,{2{p_3}+{p_4} \over 2},{m_{c}})&=&F_{13}+{\frac{1}{\varepsilon_{UV}}} \NO\\
B(0,0,{-{p_3}-{p_4} \over 2},0)&=&F_{12}+{\frac{1}{\varepsilon_{UV}}} \NO\\
B(0,{m_{c}},{{p_3}+{p_4} \over 2},{m_{c}})&=&F_{11}+{\frac{1}{\varepsilon_{UV}}} \NO\\
B(0,{m_{c}},{p_3}+{p_4},{m_{c}})&=&F_{10}+{\frac{1}{\varepsilon_{UV}}}\NO\\
B(0,0,{{p_4} \over 2},{m_{c}})&=&F_9+{\frac{1}{\varepsilon_{UV}}} 
\eea
\item Three point:
\bea
C(0,0,{{p_3} \over 2},{m_{c}},{-{p_3} \over 2},{m_{c}})&=&{\displaystyle{1 \over 
4{m_{c}}^2}}(F_8-{\frac{2\pi^2}{v}}-{\frac{2}{\varepsilon_{IR}}}) \NO\\
C(0,0,{{p_4} \over 2},{m_{c}},{2{p_3}+{p_4} \over 2},{m_{c}})&=&{\displaystyle{1
 \over 4{m_{c}}^2}}F_7 \NO\\
C(0,0,{-{p_4} \over 2},{m_{c}},{{p_3}+{p_4} \over 2},0)&=&{\displaystyle{1 \over 
4{m_{c}}^2}}F_6 \NO%\\
\eea
\bea
C(0,0,{{p_3} \over 2},{m_{c}},{{p_3}+{p_4} \over 2},0)&=&{\displaystyle{1 \over 4
{m_{c}}^2}}F_5\NO\\
C(0,0,{2{p_3}+{p_4} \over 2},{m_{c}},{{p_3}+{p_4} \over 2},0)&=&{\displaystyle{1
 \over 4{m_{c}}^2}}F_4\NO\\
C(0,0,{{p_3}+2{p_4} \over 2},{m_{c}},{-{p_3} \over 2},{m_{c}})&=&{\displaystyle{1
 \over 4{m_{c}}^2}}F_3\NO\\
C(0,0,{{p_3} \over 2},{m_{c}},-{p_4},0)&=&{\displaystyle{1 \over 4{m_{c}}^2}}F_2
\NO\\
C(0,0,{{p_4} \over 2},{m_{c}},{{p_3}+2{p_4} \over 2},{m_{c}})&=&{\displaystyle{1
 \over 4{m_{c}}^2}}F_1 
%\NO\\
\eea
\item Four point:
\bea
&&D(0,0,{2{p_3}+{p_4} \over 2},{m_{c}},{-{p_4} \over 2},{m_{c}},{{p_4} \over 2},
{m_{c}})
\NO\\
&=&{\displaystyle{1 \over 16{m_{c}}^4s}}(F_{15}s-{\frac{4\pi^2}{v}}- 
{\frac{4}{\varepsilon_{IR}}}) \NO\\[3mm]
&&D(0,0,{{p_3} \over 2},{m_{c}},{-{p_3} \over 2},{m_{c}},{-{p_3}-{p_4} \over 2},0)
\NO\\
&=&{\displaystyle{1 \over 16{m_{c}}^4s}}(F_{16}s-{\frac{8\pi^2}{v}}-{\frac{8}{\varepsilon_{IR}}})\NO\\[3mm]
&&D(0,0,{-{p_4} \over 2},{m_{c}},{2{p_3}+{p_4} \over 2},{m_{c}},{{p_3}+{p_4}
 \over 2},0)
\NO\\&=&
{\displaystyle{1 \over 16{m_{c}}^4}}F_{17}
%\NO
\eea
\item Five point:
\bea
&&E(0,0,{-{p_3} \over 2},{m_{c}},{-{p_3}-2{p_4} \over 2},{m_{c}},{{p_3} \over 2},
{m_{c}},{-{p_3}-{p_4} \over 2},0)
\NO\\&=&{\displaystyle{1 \over 64{m_{c}}^6s^2}}(F_{18}
s^2-{\frac{32\pi^2}{v}}-{\frac{32}{\varepsilon_{IR}}}) 
%\NO\\
\eea
\end{description} 
And here is the results for $F_i$ where $f_i$ and $a_i$ are defined as before.
\bea
F_1&=&{\displaystyle{1 \over s(s-4)}}4{\pi}iz_{1}(a_{6}-a_{5}+2a_{3}-4a_{1})+f_1 \NO\\
F_2&=&{\displaystyle{1 \over s(s-4)}}2{\pi}iz_{1}(-a_{5}+2a_{3}-2a_{1})+f_2 \NO\\
F_3&=&{\displaystyle{1 \over s(s-4)}}2{\pi}iz_{1}(-2a_{12}-a_{10}+a_{8}+a_{5})
+f_3 \NO%\\
\eea
\bea
F_4&=&{\displaystyle{1 \over s(s-4)}}4{\pi}iz_{1}(a_{12}-a_{3}+a_{1})+f_4 \NO\\
F_5&=&{\displaystyle{1 \over s(s-4)}}4{\pi}iz_{1}(-2a_{7}+a_{6}+2a_{1})+f_5 \NO\\
F_6&=&{\displaystyle{1 \over s(s-4)}}4{\pi}iz_{1}(2a_{13}-a_{6}-a_{5}-2a_{1})+f_6 \NO\\
F_7&=&{\displaystyle{1 \over s(s-4)}}2{\pi}iz_{1}(2a_{12}-a_{5})+f_7 \NO\\[3mm]
F_8&=&2(-2a_{1}+2) \NO\\[3mm]
F_9&=&2(a_{1}+1) \NO\\[3mm]
F_{10}&=&{\displaystyle{1 \over s}}[z_{2}(2a_{9}-a_{6}+i{\pi})+2(a_{1}s+s)] \NO\\[3mm]
F_{11}&=&{\displaystyle{1 \over s}}[z_{1}(-2a_{7}+a_{6}+2a_{1}+i{\pi})+2(a_{1}s+s
)] \NO\\[3mm]
F_{12}&=&-a_{6}+2a_{1}+i{\pi}+2 \NO\\[3mm]
F_{13}&=&{\displaystyle{1 \over 2s+1}}[2a_{1}(s+1)+2(-a_{6}s+i{\pi}s+2s+1)] \NO\\[3mm]
F_{14}&=&{\displaystyle{1 \over 4}}(2a_{1}+1) \NO\\[3mm]
F_{15}&=&{\displaystyle{1 \over s^2}}[8{\pi}i(-s+z_{2})+8z_{2}(2a_{9}-a_{6})+8s(
a_{6}+1)] \NO\\[3mm]
F_{16}&=&{\displaystyle{1 \over s}}(-32a_{1}+16) \NO\\[3mm]
F_{17}&=&{\displaystyle{1 \over s^2}}[16{\pi}i(-s+z_{2})+16z_{2}(2a_{9}-a_{6}) 
\NO\\[3mm]&&
+16(a_{6}s+2a_{1}s)] \NO\\[3mm]
F_{18}&=&{\displaystyle{1 \over s^3}}[32{\pi}i(s-2)(s-z_{2})+32z_{2}(s-2)(-2a_{9}+a_{6})
\NO\\[3mm]&&
 -32a_{6}s(s-2)-64a_{1}s^2+64s] 
\eea
\section{The scheme to treat $\gamma_5$ in D-dimension}
As we all know, there is no explicit definition for $\g_5$ in $D$ dimensions. Usually the following relations is used when one encounters $\g_5$ in $D$ dimensions:
\bea
\{\g_5,\g_\mu\}&=&0, \\
%\mathrm{Tr}(\g_5)&=&0, \\
%\mathrm{Tr}(\g_5\hat{p_1}\hat{p_2})&=&0, \\
%\mathrm{Tr}(\g_5\hat{p_1}\hat{p_2}\hat{p_3})&=&0, \\
\mathrm{Tr}(\g_5\hat{p_1}\hat{p_2}\hat{p_3}\hat{p_4})&=&4i\e_{\mu\nu\a\b}p_1^\mu p_2^\nu p_3^\alpha p_4^\beta .
\eea
Notice that $\e_{\mu\nu\a\b}$ goes zero when any of it's indices is out of the 4 dimensions.

While calculating the trace of the product of several matrices that contain $\g_5$ and indices in $D$ dimensions, different ways may lead to different results. For example, when calculating the trace of matrix $M=\g_5\g_\mu\hat{p_1}\hat{p_2}\hat{p_3}\hat{p_4}\g^\mu$, we have two different routes as shown in Fig.~\ref{fig:g5}:
\begin{figure}
\center{
\includegraphics*[scale=0.7]{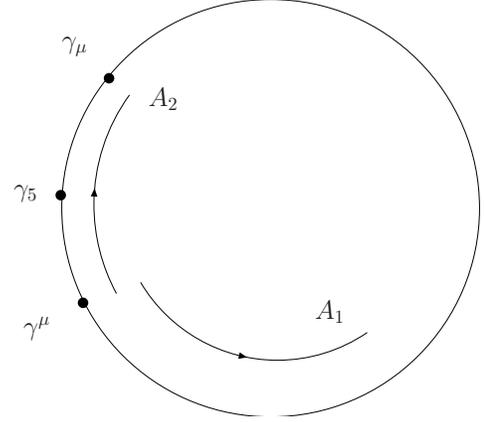}% Here is how to import EPS art
\caption{\label{fig:g5}Trace calculation including $\g_5$. }}
\end{figure}
\begin{enumerate}
\item In route $A_1$, the summing up of the index $\mu$ does not go across $\g_5$.
\bea
\mathrm{Tr}(M)|_{A_1}&=&\mathrm{Tr}\biggl\{\g_5\biggl[(D-2)\hat{p_1}\hat{p_2}\hat{p_3}\hat{p_4} +2\hat{p_2}\hat{p_1}\hat{p_3}\hat{p_4}\NO\\[3mm]
&& -2\hat{p_3}\hat{p_1}\hat{p_2}\hat{p_4} +2\hat{p_4}\hat{p_1}\hat{p_2}\hat{p_3}\biggr]\biggr\} \NO\\[3mm]
&=&4i(D-8)\e_{\mu\nu\a\b}p_1^\mu p_2^\nu p_3^\a p_4^\b .
\eea
\item In route $A_2$, it does go across $\g_5$.
\bea
\mathrm{Tr}(M)|_{A_2}&=&-\mathrm{Tr}\biggl(\g_5\g^\mu\g_\mu\hat{p_1}\hat{p_2}\hat{p_3}\hat{p_4}\biggr) \NO\\[3mm]
&=&-D\mathrm{Tr}\biggl(\g_5\hat{p_1}\hat{p_2}\hat{p_3}\hat{p_4}\biggr) \NO\\[3mm]
&=&-4iD\e_{\mu\nu\a\b}p_1^\mu p_2^\nu p_3^\a p_4^\b .
\eea
\end{enumerate} 
It is easy to find that $\mathrm{Tr}(M)|_{A_1}=\mathrm{Tr}(M)|_{A_2}$ in 4 dimensions, but in $D$ dimensions they're different to each other. So we should always take same route when dealing with traces containing $\g_5$ in order to keep our final finite result consistent. In FDC, route $A_1$ is taken.
%\newpage %Just because of unusual number of tables stacked at end
\bibliography{twain.bib}% Produces the bibliography via BibTeX.
\end{document}